 \documentclass[preprint,preprintnumbers,amsmath,amssymb]{revtex4}
 \usepackage{amsmath,amssymb,bm,epsfig}
 \usepackage{color}
 \usepackage{natbib}
 \usepackage{hyperref}
 \usepackage{ulem}
 \usepackage{graphicx}

\def \beq{\begin{equation}}
\def \eeq{\end{equation}}
\def \beqa{\begin{eqnarray}}
\def \eeqa{\end{eqnarray}}
\def \la{\langle}
\def \ra{\rangle}
\def \l{\left(}
\def \r{\right)}

\newcommand{\nn}{\nonumber}

\usepackage{xspace}

\begin{document}
\title{An estimate of the bulk viscosity of the hadronic medium
}

\author{Golam Sarwar}
\email{golamsarwar@vecc.gov.in}
\affiliation{Theoretical Physics Division, 
Variable Energy Cyclotron Centre, 1/AF Bidhannagar, 
Kolkata, 700064, India}

\author{Sandeep Chatterjee}
\email{sandeepc@vecc.gov.in}
\affiliation{Theoretical Physics Division, 
Variable Energy Cyclotron Centre, 1/AF Bidhannagar, 
Kolkata, 700064, India}

\author{Jane Alam}
\email{jane@vecc.gov.in}
\affiliation{Theoretical Physics Division, 
Variable Energy Cyclotron Centre, 1/AF Bidhannagar, 
Kolkata, 700064, India}

\begin{abstract}
The bulk viscosity ($\zeta$) of the hadronic medium has been estimated within the ambit of the  
Hadron Resonance Gas (HRG)  model including the Hagedorn density of states. 
The HRG thermodynamics within a grand canonical ensemble provides the mean hadron 
number as well as its fluctuation. The fluctuation in the chemical composition of the hadronic medium 
in the grand canonical ensemble can result in non-zero divergence of the hadronic 
fluid flow velocity, allowing us to estimate the $\zeta$ of the hadronic matter upto a relaxation 
time. We study the influence of the hadronic spectrum on $\zeta$ and find its correlation with 
the conformal symmetry breaking (CSB) measure, $\epsilon-3P$. We estimate $\zeta$ along 
the contours with constant, $S/N_B$  (total entropy/net baryon number)  in the 
$T-\mu$ plane (temperature-baryonic chemical potential) for $S/N_B=30, 45$ and 300. 
We also assess the value of $\zeta$ on the chemical freezeout curve for various centre 
of mass energy ($\sqrt{s_{NN}}$) and find that  
the bulk viscosity to entropy density ratio, $\zeta/s$ is larger at FAIR than LHC energies. 

PACS numbers:12.38.Mh,24.10.Pa,24.85.+p,25.75.Dw

\end{abstract}
\maketitle

\section{Introduction}\label{sec.intro}
Hydrodynamics is the study of the slowly varying degrees of
freedom of the system involving continuity equations of the conserved charges of the underlying
microscopic interactions.
Viscous relativistic hydrodynamics has been a very successful 
framework to describe the evolution of 
the fireball created in heavy ion collisions at relativistic energies (HICRE)
with a few free parameters which are extracted by fits to
data~\cite{Heinz:2013th,Gale:2013da}. 
The following equations along with the initial conditions and
equation of state (EoS) govern the  progression of the matter 
produced in HICRE,  
\beqa
\partial_\mu T^{\mu\nu}&=&0\\\label{eq.tmunu}
\partial_\mu \mathcal{J}_\kappa^\mu&=&0\label{eq.jmu}
\eeqa
where $T^{\mu\nu} (\mu,\,\nu=0,1,2,3)$ is the energy momentum tensor and $\{\mathcal{J}_\kappa^\mu\}$ refers to the  
current density corresponding to the conserved charges, ({\it e.g}, $\kappa\equiv$ net baryon number, net electric charge, etc.) 
$T^{ij} (i\,,j=1,2,3)$ for a non-viscous hydrodynamical system in the local rest frame is given by,
\begin{eqnarray}
T_{ij}&=&P\delta_{ij}
\label{eq.tmununovisc}
\end{eqnarray}
where $P$ is the isotropic thermodynamic pressure of the system. In case of Navier Stokes viscous hydrodynamics, 
the system's response to the gradients of the fluid flow four velocity $u^\mu$.
Upto first order in derivatives of $u^\mu$ two such transport coefficients, namely shear $(\eta)$ 
and bulk $(\zeta)$ viscosities appear as,
\begin{eqnarray}
T_{ij}&=&P\delta_{ij}+\eta\l\frac{\partial u_i}{\partial x^j} + \frac{\partial u_j}{\partial x^i} - 
\frac{2}{3}\delta_{ij}\nabla\cdot u\r + \zeta\delta_{ij}\nabla\cdot u
\label{eq.tmunuvisc}
\end{eqnarray}
That is the leading order correction ($\delta T_{ij}$) to the energy momentum tensor 
due to dissipation is introduced through the shear ($\eta$) and bulk ($\zeta$)  viscous coefficients  
such that $T_{ij}=P\delta_{ij}+\delta T_{ij}$. 
The effects of these transport coefficients on the hydrodynamic evolution of the strongly interacting 
fireball and on various observables in HICRE,  {\it e.g.}  the elliptic flow, spectra of hadronic and electromagnetically 
interacting particles, etc 
have been studied extensively~\cite{Romatschke:2007mq,Song:2012ua,Alver:2010gr,Gardim:2011xv,Niemi:2012aj,dusling}. The general 
consensus reached is that the matter created in heavy ion collisions behaves almost like a perfect liquid 
~\cite{Gyulassy:2004zy,Arsene:2004fa,Adcox:2004mh,Back:2004je,Adams:2005dq} 
with $\eta/s$ close to the KSS (Kovtun, Son and Starinets) bound~\cite{Kovtun:2004de}. On the other hand, 
the issue of bulk viscosity is far from settled. In the earlier works the contribution of $\zeta$ was neglected. 
Inspired by the AdS/CFT correspondence, the ${\cal N}=4$ supersymmetric Yang-Mills theory has been
used by several authors to estimate the
shear viscosity of the QGP and other strongly correlated system~\cite{Kovtun:2004de}. 
The $\zeta/s$  is always zero in this approach.  However,
it is shown in~\cite{gubser} that $\zeta/s\sim 0.1$ for certain classes of black hole solutions.   
Moreover,  the lattice QCD (LQCD) based studies~\cite{Meyer:2007dy,Karsch:2007jc} has indicated
that the bulk viscosity 
could be as large  as shear viscosity at the vicinity of the QCD phase transition. Similar conclusions have also 
been drawn from  calculations done by using QCD inspired effective models~\cite{model1,model2,model3}. 
Thus, there have been studies 
where $\zeta$ was included into hydrodynamic simulations. It was found that $\zeta$ affects  
the low momentum hadron spectra as well as the elliptic flow significantly~\cite{ryu}. 

The phenomenological relevance of $\zeta$ has fuelled efforts to estimate it by using various models. In this work we 
provide an estimate of $\zeta$ within the ambit of HRG model which has been quite successful in 
describing the low temperature QCD thermodynamics. Lately, bulk viscosity of the hadronic medium has been 
computed in various 
schemes~\cite{Lu,sukanya,juan,guru,purnendu,wang,sasaki,noronha}. 
In this work we intend to study the role played by the phase space in deciding the bulk viscosity of the 
hadronic medium. 

The paper is organized as follows. In the next section we outline the formalism used to estimate
the bulk viscosity of hadronic system. We discuss the results of the present work in section III
and devote section IV for summary and discussions.   

\section{Formalism}\label{sec.formalism}
A fluid in equilibrium can fluctuate to a non-equilibrium state in many ways. Depending 
on the mode of fluctuation there is an onset of the corresponding dissipative process to 
counter this fluctuation for  maintaining the equilibrium. Within the ambit of linear 
response theory, the medium response allows us to compute the transport coefficients like 
$\eta$, $\zeta$ etc. As seen in Eq.~\ref{eq.tmunuvisc}, $\eta$ 
is connected to the traceless part of $T_{ij}$ given by Eq.~\ref{eq.tmunuvisc} 
and $\zeta$ is related to the trace of $T_{ij}$ for a compressible fluid warned
by the presence of $\nabla\cdot u$ which is related to the rate of change of volume ($V$)
associated with the uniform expansion or compression through the continuity equation:
\beq
\frac{\partial n}{\partial t}+\vec{\nabla}\cdot\,(n\vec{u})=0
\label{conteq}
\eeq
substituting the density $n=N/V$ in Eq.~\ref{conteq} ($N$ is the total number and $V$ is the volume),
one obtains $\nabla\cdot u=V^{-1}dV/dt$~\cite{Lu}. This also indicates that the bulk viscosity will
vanish for an incompressible fluid.
The deviation in the pressure due to change in the bulk not followed from the equation of state (EoS)
can be connected to the bulk viscosity as follows: 
\beq
\delta P = \zeta\nabla\cdot u
\label{eq.deltaP}
\eeq

In this work we will consider 
the situation where such flow field arises due to change in hadron yield from the equilibrium number.
For a single component hadron gas assuming adiabaticity, it has been shown in ~\cite{Paech:2006st} that:
\beqa
\delta P &=& {\l\frac{\partial P}{\partial n}\r}_{\epsilon}\frac{\partial n}{\partial s}s\nabla\cdot u \tau_R
\label{eq.deltapsingle}
\eeqa
where $\tau_R$ is the relaxation time scale of the system, which is the inverse of
the rate of number changing process responsible to maintain chemical equlibrium. 
Now within the HRG formalism, the total pressure $P$ is given by sum over the
partial pressure, $P_i$  due to each hadron species, $i$,
\beqa
P&=&\sum P_i
\label{eq.pressuretotal}
\eeqa
Thus the fluctuation in the total pressure $P$ can be expressed as
\beqa
\delta P &=& \sum_i\delta P_i = \left[\sum_i {\l\frac{\partial P_i}{\partial n_i}\r}_{\epsilon_i}
\frac{\partial n_i}{\partial s_i}s_i\tau_R^i\right]\nabla\cdot u
\label{eq.deltaPtherm}
\eeqa
$\tau_R^i$ is the relaxation time of the species, $i$. 
The fluid flow velocity field, $u^{i}$ 
being a hydrodynamic variable is same for all the hadron species. On comparing Eqs.~\ref{eq.deltaP} 
and \ref{eq.deltaPtherm}, we find the expression for $\zeta$
\beq
\zeta = \sum_i {\l\frac{\partial P_i}{\partial n_i}\r}_{\epsilon_i}
\frac{\partial n_i}{\partial s_i}s_i\tau_R^i \label{eq.zeta} \\
\eeq
This relation can be used to derive the expression for bulk viscosity (see appendix for details) as,
\beqa
\zeta&=& \sum_i \left[\frac{\l\frac{\partial P_i}{\partial T}\r\l\frac{\partial\epsilon_i}{\partial\mu_i}\r-
\l\frac{\partial P_i}{\partial \mu_i}\r\l\frac{\partial\epsilon_i}{\partial T}\r
}{\l\frac{\partial n_i}{\partial T}\r\l\frac{\partial\epsilon_i}{\partial\mu_i}\r
-\l\frac{\partial n_i}{\partial\mu_i}\r\l\frac{\partial\epsilon_i}{\partial T}\r}\right]
\left[\l\frac{\partial n_i}{\partial T}\r\l\frac{\partial T}{\partial s_i}\r+
 \l\frac{\partial n_i}{\partial\mu_i}\r\l\frac{\partial\mu_i}{\partial s_i}\r\right] s_i\tau_R^i
\label{eq.zetaix}
\eeqa
In order to estimate $\tau_R^i$ we need to
know the cross sections of all the possible processes through which hadron, $i$ interacts 
with all the hadrons and  resonances. As all these required cross sections  are not
known presently and we are interested in studying the effects of phase space on $\zeta$,
a constant cross sections for all the hadronic processes is assumed  
in the spirit of Ref.~\cite{denicolprc88} and treat the
relaxation time as a constant to write down the ratio $\zeta/\tau_R$ as:
\beqa
\frac{\zeta}{\tau_R}&=& \sum_i \left[\frac{\l\frac{\partial P_i}{\partial T}\r\l\frac{\partial\epsilon_i}{\partial\mu_i}\r-
\l\frac{\partial P_i}{\partial \mu_i}\r\l\frac{\partial\epsilon_i}{\partial T}\r
}{\l\frac{\partial n_i}{\partial T}\r\l\frac{\partial\epsilon_i}{\partial\mu_i}\r
-\l\frac{\partial n_i}{\partial\mu_i}\r\l\frac{\partial\epsilon_i}{\partial T}\r}\right]
\left[ \l\frac{\partial n_i}{\partial T}\r\l\frac{\partial T}{\partial s_i}\r+
\l\frac{\partial n_i}{\partial\mu_i}\r\l\frac{\partial\mu_i}{\partial s_i}\r\right] s_i
\label{eq.zetai}
\eeqa
Eq.~\ref{eq.zetai} has been used  to estimate the $\zeta$ for HRG in this work. 
Each term in Eq.~\ref{eq.zetai} can be computed from $P_i$ and its derivatives, where $P_i$ is given by,
\beqa
P_i &=& \frac{T}{V}\ln\, Z_i\l T,V,\mu_i\r\nn\\
&=&\sum_i \frac{a g_i}{2\pi^2}T^4\int_0^{\infty}dxx^2
\ln\left[1+a\exp\left[-\l\sqrt{x^2+\l\frac{m_i}{T}\r^2}-\frac{\mu_i}{T}\r\right]\right]
\label{eq.Pi}
\eeqa
where $a=-1$ for mesons (Bosons) and $+1$ for baryons (Fermions). Consequently the corresponding entropy density ($s_i$), 
number density ($n_i$) and energy density ($\epsilon_i$) are given by, 
\beq
s_i = \frac{\partial P_i}{\partial T},\,\,\, n_i = \frac{\partial P_i}{\partial \mu_i},\,\,\,\,
\epsilon_i = T\frac{\partial P_i}{\partial T}-P_i+\mu_i\frac{\partial P_i}{\partial\mu_i}
\label{eq.ni}
\eeq
It can be easily checked from Eqs.~\ref{eq.zetai}, \ref{eq.Pi} and \ref{eq.ni} that when the  
hadron, $i$  is massless with $\epsilon_i=3P_i$ then $\zeta_i$ vanishes as 
$\l\frac{\partial P_i}{\partial T}\r\l\frac{\partial\epsilon_i}{\partial\mu_i}\r-
\l\frac{\partial P_i}{\partial \mu_i}\r\l\frac{\partial\epsilon_i}{\partial T}\r=0$. 
Eqs. \ref{eq.Pi} and \ref{eq.ni} can be used to reproduce the known thermodynamic
expressions for pressure, entropy density, number density, energy density, etc., 
both for relativistic and non-relativistic limits. In turn these quantities 
can be used in Eq.~\ref{eq.zetai} to estimate the bulk viscosity to relaxation
time ratio.

\section{Results}
\label{sec.results}
The bulk viscosity, $\zeta$ for a HRG system can be calculated by using 
Eq.~\ref{eq.zetai} where all the particles as 
listed in the Particle Data Book~\cite{Agashe:2014kda} of mass upto $2.5$ GeV are included. In order to understand the results 
for the full HRG, we first investigate a system with single species of mass, $m_l$ and 
then the other with two different species of  masses $m_l$ and $m_h$ with $m_h>m_l$.  
We study the interplay of the two different mass scales on the  temperature dependence of   $\zeta$ and its 
correlation with the CSB (conformal symmetry breaking) measure, $\Delta=\epsilon-3P$~\cite{fernandez}.
\begin{figure}
\centerline{\includegraphics[height=60mm, width=60mm]{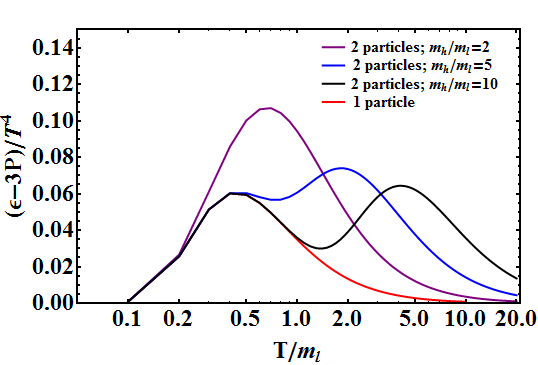}}
\caption{Variation of $\l\epsilon-3P\r/T^4$ with $T/m_l$ for different values 
   of the ratio of the masses of the heavy to light particle, $m_h/m_l$ for vanishing  
   chemical potential.}
   \label{fig1}
\end{figure}

\begin{figure}
\centerline{\includegraphics[height=60mm, width=60mm]{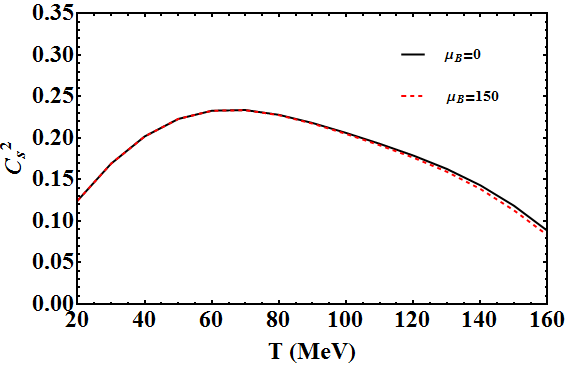}}
   \caption{Variation of  the square of the speed of sound with $T$ for 
   zero and non-zero net baryon density. 
   }
   \label{cs2}
\end{figure}

\begin{figure}
\centerline{\includegraphics[height=60mm, width=60mm]{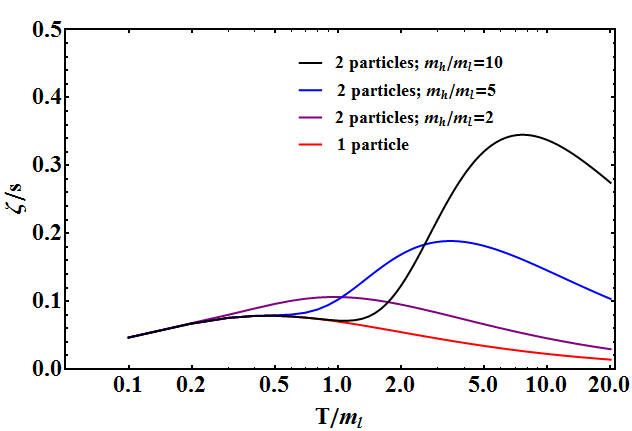}}
   \caption{Variation of $\zeta/s$ with $T/m_l$ for different values 
   of the ratio of the masses of the heavy to light particle, $m_h/m_l$ for vanishing  
   chemical potential.}
   \label{fig2}
\end{figure}
We have plotted  $\Delta/T^4$ as function of $T$  in Fig.~\ref{fig1} for
systems with different composition and masses to elucidate the role of hadronic masses in 
$\Delta$ and subsequently  in bulk viscosity (Fig.\ref{fig2}).
The curves in ~Fig.\ref{fig1} stand for different values of the 
ratio, $m_h/m_l$. The qualitative features of the plots remain same when we replace 
the bosons by fermions. 
Results displayed in Fig.~\ref{fig1} indicate that $\Delta/T^4\,\rightarrow 0$
both for the non-relativistic $m/T>>1$ and massless limits 
$m/T\rightarrow 0$.  To understand the variation  of $\Delta/T^4$
with $T/m$, first consider a system at temperature $T$ with single species of mass $m_l$.  
In the high temperature limit the
pressure-energy density relation becomes $P=\epsilon/3$ giving rise to $\Delta = 0$. 
In the limit of large  $m/T$ (small $T/m$) the CSB measure varies as: $\Delta\sim e^{-m/T}$,
becomes vanishingly small. That is for both small and large $T$, $\Delta\rightarrow 0$, with 
an intermediary peak at $T/m_l\sim 0.5$. 

Now we consider a two-particle system with  masses $m_l$ and $m_h$ ($m_h>m_l$). 
First consider the case with $m_h/m_l=\infty$. The heavier particle 
does not contribute to the thermodynamics. Hence this is essentially a single particle system. We find a 
single peak around $T\sim 0.5m_l$. Next, we plot for the case with $m_h/m_l=10$. The large 
separation in the masses of the two particles results in distinct two peaks at $T/m_l\sim0.5$ and
$T/m_l\sim 5$ (i.e. at $T/m_h\sim 0.5$).  
For $m_h/m_l=5$, similar structure is found with closer peaks
and reduced dip between the two peaks. We observe that the peak associated with
the lighter particle has converted to a shoulder-like structure. 
Finally for $m_h/m_l=2$, the peaks have partially merged and we are left with 
only a single (broader) peak at $T\sim0.5\l0.5m_l+0.5m_h\r=0.75m_l$. 

The presence of the massive hadrons 
does not allow  the system to satisfy the relation $\epsilon=3P$ 
{\it i.e.} the conformal symmetry is broken for the entire range of $T$ 
both for zero and non-zero $\mu_B$. This is evident from the estimation of the 
speed of sound ($c_s^2$) which  remains below $1/\sqrt{3}$ (Fig.~\ref{cs2}) for
the entire $T$ range considered.

It is expected that the temperature and mass dependences of CSB discussed above will be reflected on $\zeta/s$
as these quantities are correlated~\cite{mcheng}. For demonstrating the 
phase space dependence of $\zeta$ we assume $\tau_R\sim 1$ fm/c. The results are depicted in 
Fig.~\ref{fig2}.  The peaks corresponding to the single and two particles systems (with $m_h/m_l=2$) 
get blurred. For $m_h/m_l=5$ and $10$ the peaks at lower $T/m_l$  get smeared, however,
for higher $T/m_l$ the peaks become broader but distinctly visible. 
In summary, the  $\zeta$ and $\Delta$ have similar $T$-variation with broader peaks in the later quantity.
For a system with many particles the $\zeta$ will be a superposition of results obtained for each of the different 
hadrons with their respective masses.  
We use Eqs.~\ref{eq.zetai}, \ref{eq.Pi} and \ref{eq.ni} to estimate the ratio $\zeta/\tau_R$ 
for a system of single particles with mass $m$ in the limits of $m/T\rightarrow 0$ 
and $m/T\rightarrow \infty$.  We find that $\zeta/\tau_R\,\sim (m/T)^2$ for
$m/T<<1$ and $\zeta/\tau_R\,\sim e^{-(m-\mu)/T}$ 
for $m/T>>1$, {\it i.e.} the bulk viscosity vanishes both
in the relativistic and non-relativistic limits - a well known result in
the literature.
  
Now we turn our attention to the HRG system. The study of HRG  is 
important because lattice QCD results indicate that at lower temperatures, the
HRG  is a good proxy for the effective degrees of
freedom of the strongly interacting matter. Therefore, it will be very useful 
to study the properties of HRG if it  is away
from equilibrium. We estimate the bulk viscosity of the HRG when it is
slightly away  from equilibrium - a situation may be confronted
during the evolution of matter formed in nuclear collisions
at relativistic energies.

The lightest hadron is the pion with $m_\pi\sim140$ MeV and 
the next hadron (kaon) is heavier by about 350 MeV. 
The hadronic degrees of freedom are expected to survive upto 
$T\sim 150-160$ MeV. Thus, in this temperature domain, $m/T>>1$ for all hadrons except pion.  
This implies that for the full HRG system, we should expect to see features 
qualitatively similar to the non-relativistic end of the plots in Fig.~\ref{fig2} {\it i.e.} 
we should see an increasing trend of $\zeta$ with $T$ for constant $\tau_R$. 

\begin{figure}
\centerline{\includegraphics[height=60mm, width=60mm]{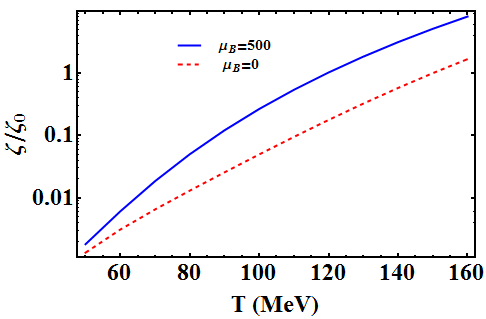}}
   \caption{The temperature variation of  $\zeta(T)/\zeta_0$
is shown here with  $\zeta_0=\zeta(T=150)$ MeV. 
}
\label{fig3}
\end{figure}

\begin{figure}
\centerline{\includegraphics[height=60mm, width=60mm]{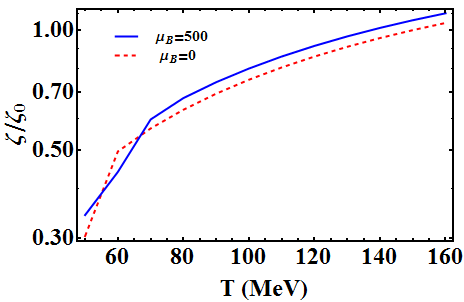}}
   \caption{Depicts the temperature variation of the bulk viscosity
    to entropy density ratio normalized to the value of the ratio at $T=150$ MeV (see text)
with constant relaxation time.}
   \label{fig4}
\end{figure}

In Figs.~\ref{fig3} and ~\ref{fig4} we have displayed the temperature variation of $\zeta/\zeta_0$
and $R_\zeta=(\zeta/s)/(\zeta_0/s_0)$ 
$[\zeta_0=\zeta(T=150$ MeV) $s_0=s(T=150$ MeV)$]$ respectively for different values 
of baryonic chemical potential,  $\mu$ within the framework of HRG model.
Please note that in the  
ratios of $\zeta$ the effects of the constant relaxation time get cancelled. 
We find that the bulk viscosity increases with both temperature and baryonic density as
expected from the discussions above. 


The results displayed so far may be improved by the following two considerations:
(i) by making the $\tau_R$ a $T$ and $\mu$ dependent quantity. We have taken constant $\tau_R$ so far, however,  
the relaxation time $\tau_R$ should depend on the thermodynamic state of the matter, {\it i.e.} 
it should vary with $T$ and $\mu$. 
To get the $T$ and $\mu$ dependence of $\tau_R$ we can use  the relation, 
\beq
\tau_R^j =\frac{1}{\sum_i\sigma_{ij}n_i\frac{\la p_i\ra}{\la E_i\ra}}
\label{eq.relaxation}
\eeq
However, as mentioned above we will assume constant cross section {\it i.e.} $\sigma_{ij}=\sigma$ and
a single relaxation time for the system, {\it $\tau_R^j=\tau_R$}. 
We find that the relaxation time,
$\tau_R$ reduces with the $T$ and $\mu$  i.e. the hotter 
and denser systems relax faster.   

(ii) By including the Hagedorn density of states (HDS) ~\cite{chatterjee, Lo}  
for counting the resonances at higher temperatures in estimating the bulk viscosity~\cite{tawfikbulk}. For this 
purpose we have used the following mass spectrum in evaluating relevant thermodynamic quantities:
\beq
\rho\l m\r = \sum_id_i\delta\l m-m_i\r+\frac{a_0}{\l m^2+m_0^2\r^{5/2}}e^{m/T_H}\label{eq.Hag}
\eeq
where the first part is the standard discrete contribution from all the PDG resonances while 
the second part is the additional contribution from the continuous HDS. $d_i=2S_i+1$ is 
the degeneracy due to the spin of the $i$th hadron with mass $m_i$. $a_0$, $m_0$ and $T_H$ are parameters extracted 
from fits of the HDS to the observed spectrum. Here we have used $a_0=0.744$ GeV$^{3/2}$, $m_0=0.529$ 
GeV and $T_H=180$ MeV as in~\cite{Lo}. The thermodynamic quantities like the energy density ($\epsilon$) may
be calculated by using the formula: $\epsilon=\int dm\rho(m)\int\frac{d^3p}{(2\pi)^3}\sqrt{p^2+m^2}f(p)$,
where $f(p)$ is the appropriate thermal distribution for Bosons or Fermions.   

\begin{figure}
\centerline{\includegraphics[height=60mm, width=60mm]{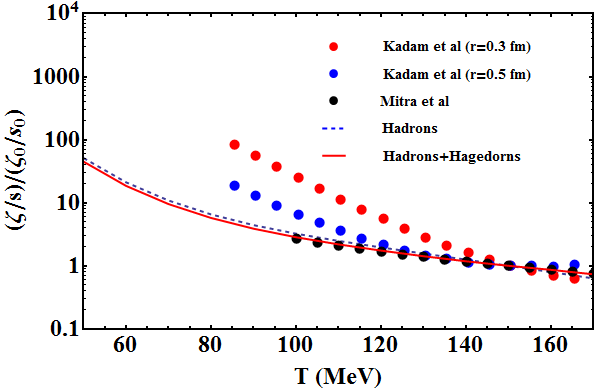}}
\caption{Variation of $R_\zeta$ (see text) as a function of $T$  for hadronic
resonances upto mass 2.5 GeV with (red line)  and without (blue dashed line) HDS 
including $T$ dependent relaxation time estimated by using Eq.~\ref{eq.relaxation}. 
We have also displayed the same quantity as obtained in other works~\cite{sukanya,guru}.}
\label{fig5}
\end{figure}

With the inclusion of these two effects as described above the temperature dependence of $R_\zeta$, 
the normalized bulk viscosity  has been evaluated and the result is displayed 
in Fig.~\ref{fig5}. The  $T$ and $\mu$ dependence of $\tau_R$ has changed the results 
both quantitatively and qualitatively. 
In sharp contrast to the results displayed in Fig.~\ref{fig4}
the ratio decreases with temperature as observed also in Refs.~\cite{Lu,sukanya}. However,
we recall that the results depicted in Fig.\ref{fig5} contains temperature 
dependent $\tau_R$ which decreases with $T$ as: $\tau_R^{-1}\propto T^2e^{-m/T}$ (due 
to $T$ dependent density and average velocity) for constant 
cross sections as assumed here. This temperature variation seems to be stronger 
than the $T$ dependent growth of the right hand side (rhs) of Eq.\ref{eq.zetai}.
As a result the ratio, $R_\zeta$ which is a product of these two factors - rhs of Eq.~\ref{eq.zetai}
and $\tau_R$ decreases with $T$.    
The outcome of the present work has particularly 
been compared with the results obtained by solving Boltzmann equation - 
in (a) relaxation time approximation with excluded volume effects in HRG ~\cite{guru}
and (b) Chapman-Enskog approximations for interacting pion gas~\cite{sukanya}. At higher 
temperature all the results converge. At lower temperature the results from
different models tend to differ. However, the  outcome of the 
present work with the inclusion of HDS (solid line)
agrees well with the results of Ref.~\cite{sukanya}.

\begin{figure}
\centerline{\includegraphics[height=60mm, width=60mm]{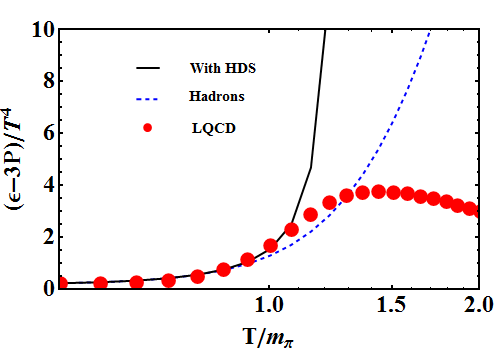}}
   \caption{The variation $(\epsilon-3P)/T^4$ with 
$T/m_\pi$ with (solid line) and without (dashed line) HDS. A comparison has been made with 
continuum extrapolated (2+1) lattice QCD data with physical quark masses~\cite{LQCD-data}.
}
   \label{fig6}
\end{figure}

In Fig.\ref{fig6} the variation of CSB measure with temperature is displayed
for HRG system with (solid line)  and without (dashed line) HDS. A significant enhancement of
CSB is observed for temperature above pion mass, more so for the case where HDS are used 
in addition to the standard PDG hadrons. This is expected as the inclusion of additional Hagedorn 
resonances result in a stronger breaking of the conformal symmetry. We have compared the CSB measure 
obtained in the HRG model with that of LQCD data and find good agreement upto about $T\sim1.1m_{\pi}$ 
which is the region of interest here. It is expected that the observed variation of CSB with $T$ will 
also reflect in the $T$ dependence of $\zeta$.

\begin{figure}
\centerline{\includegraphics[height=60mm, width=60mm]{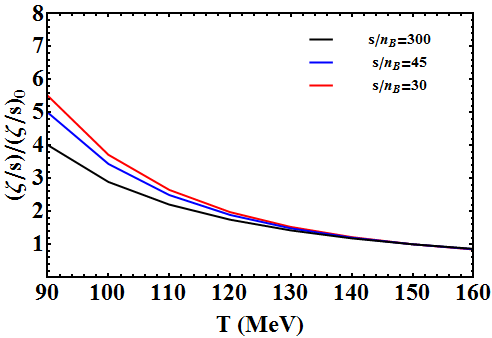}}
\caption{The variation of  bulk viscosity to entropy density ratio (normalized at $T=150$ MeV) 
with  temperature along constant $S/N_B$ contours for $S/N_B=30, 45$ and 300.  
}
\label{fig7}
\end{figure}

In HICRE, for a given $\sqrt{s_{NN}}$ a hot and dense medium
will be created  with entropy, $S$ and net baryon number, $N_B$.
For an isentropic  expansion of the system, the $S/N_B$ ratio will be constant
throughout the evolution. Therefore, the system will evolve along a trajectory in
the $T-\mu$ plane corresponding to a constant $S/N_B$ contour.  
We use this evidence to evaluate $\zeta$ along the constant $S/N_B$  contours in
the $T-\mu$ plane for  $S/N_B=$30, 45 and 300.
These values of $S/N_B$ may correspond to AGS (FAIR), SPS and RHIC collision conditions~\cite{ejiri}
(see also ~\cite{tawfikcomments}).
The variation of $R_\zeta$ with $T$ along these contours
are depicted in Fig.~\ref{fig7}.  The relaxation
time is estimated by using Eq.~\ref{eq.relaxation}  with the value of  cross section, 
$\sigma=\pi$ fm$^2$  as in~\cite{denicolprc88}. It is observed that 
the magnitude of $\zeta/s$ at low temperature domain is higher for $S/N_B=30$ (corresponds to higher $\mu$)
compared to $S/N_B=300$.  At higher temperatures the values of bulk viscosity 
seems to converge for all the values of $S/N_B$.   

\begin{figure}
\centerline{\includegraphics[height=60mm, width=60mm]{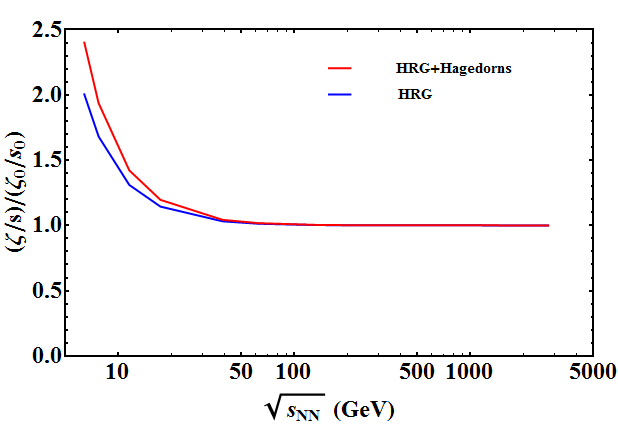}}
   \caption{The variation of  $R_\zeta$ as a function of $\sqrt{s_{NN}}$ with (red line)  and without (blue line) HDS. 
}
   \label{fig8}
\end{figure}

In HICRE the centre of mass  energy ($\sqrt{s_{NN}}$) can be connected  to
the values of $T$ and $\mu$ at the chemical freeze-out curve by analyzing the hadronic 
yields~\cite{cleymans1,cleymans2} as follows. In Ref.~\cite{cleymans2} the chemical freeze-out  curve
has been parametrized as: $T(\mu)=0.166-0.139\mu^2-0.053\mu^4$ and the $\sqrt{s_{NN}}$ 
dependence of $\mu$ has been fitted with $\mu=1.308(1+0.273\sqrt{s_{NN}})^{-1}$. 
Using these parameterizations the normalized bulk viscosity has been estimated and the results
have been displayed as a function of  $\sqrt{s_{NN}}$ in Fig~\ref{fig8}   
with (dashed line) and without (solid) incorporating the HDS.
At high $\sqrt{s_{NN}}$ the inclusion of Hagedorn spectra does not make any difference, however, 
at lower values of $\sqrt{s_{NN}}$  Hagedorn spectra enhances the bulk viscosity. 
The important point to be noted here is that at lower beam energy the bulk 
viscosity is larger (for larger $\mu$).
Therefore, $\zeta$ will play a more important role at FAIR than LHC experiments.


\section{Summary and discussions}\label{sec.summary}
Using grand canonical ensemble the bulk viscosity of the hadronic medium 
has been estimated within the ambit of the HRG model approach.
The grand canonical ensemble of HRG provides the mean hadron 
number as well as the fluctuations in the chemical composition of the hadronic medium.
These fluctuations grant a non-zero divergence for the hadronic 
fluid flow velocity, offering an opportunity to evaluate the hadronic 
bulk viscosity $\zeta$ upto a relaxation time. First  we have considered both 
single and two hadronic systems with different masses to exemplify the role of 
hadronic masses on the CSB and bulk viscosity.  Then we proceed to evaluate the  $\zeta$ for HRG model and
eventually include the HDS in the  calculations. 
We find that the inclusion of HDS enhances the bulk viscosity of the system at lower $\sqrt{s_{NN}}$.
We would like to note here that recently it has been shown that a considerable improvement 
in the HRG framework in describing LQCD data is obtained by simultaneous inclusion of the HDS as well as 
a hard core repulsion between the hadrons within an excluded volume approach~\cite{goren}. 
In the low temperature domain ($T\simeq 150$ MeV) it is found that the pressure and  energy density 
estimated without excluded volume effect remain within error bars of the lattice QCD results~\cite{goren}. 
However, the agreement with the energy density turns to be better at higher $T$ with excluded volume effect.    
We leave this interesting exercise about the simultaneous role of the finite size of the hadrons as well 
as the HDS on the bulk viscosity of the hadronic medium for the future.
We also estimate $\zeta$ along the constant $S/N_B$ contours and find that $\zeta/s$ is enhanced for
lower $S/N_B$.  $\zeta/s$ has also been evaluated along the chemical freeze-out curve obtained from the parameterization
of hadronic yields~\cite{cleymans2}  and found that the $\zeta/s$ is larger at FAIR than LHC energy region.
This indicates that the  bulk viscosity will play more crucial role in nuclear collision at FAIR than LHC energies.  

A few words on the $T$ and $\mu$ dependence of the 
bulk viscosity arising from the phase space factors are in order here. 
In the present work we have assumed a constant $\sigma$ for all the hadronic processes, however, in reality
the situation could be more complex with $T$ and $\mu$ dependent cross sections
to be considered for all types of possible reactions undergoing in the medium.
As mentioned earlier the cross sections for the hadronic reactions involving all the
resonances and Hagedorn states are not known presently. Therefore, we have assumed a constant cross section
~\cite{denicolprc88} and demonstrated the $T$ and $\mu$ dependence of 
the bulk viscosity originating from the phase space factors only. The lack of these cross sections
does not allow us to estimate the relaxation time from microscopic interactions. Therefore, we assume
that the  relaxation time of a hadron $h$ is 
$\tau_h=\tau_R \pm \delta_h$, 
where $\tau_R$ is the average relaxation time scale of the full system and $\delta_h$ is the deviation of $\tau_h$ from it.
In this work, it is assumed that $\tau_R >> \delta_h$ for all $h$.

%
%
%
Acknowledgement: SC  \& GS thank DAE for financial support.
SC is supported by CNT project: PIC XII-R\& D - VECC - 5.02.0500.

\section{Appendix}
\label{app}
In this appendix we will show in details the computation of the bulk viscosity. For 
simplicity we  assume that the state of the hadronic matter
under consideration can be described by two independent thermodynamic variables:
{\it i.e.} temperature ($T$) and baryonic chemical potential ($\mu$).
When we take partial derivative w.r.t. $T$ it is understood that $\mu$ is constant and vice-versa 
and hence we do not mention this explicitly. The differential of $P\l T,\mu\r$ can be written as
\beqa
dP &=& \l\frac{\partial P}{\partial T}\r dT + \l\frac{\partial P}{\partial \mu}\r d\mu\nn\\
\frac{\partial P}{\partial n} &=& \l\frac{\partial P}{\partial T}\r \frac{\partial T}{\partial n} + 
\l\frac{\partial P}{\partial \mu}\r \frac{\partial \mu}{\partial n}\nn\\
\l\frac{\partial P}{\partial n}\r_{\epsilon} &=& \l\frac{\partial P}{\partial T}\r \l\frac{\partial T}
{\partial n}\r_\epsilon + \l\frac{\partial P}{\partial \mu}\r \l\frac{\partial \mu}{\partial n}\r_\epsilon\nn
\eeqa
 where  $\l\frac{\partial n}{\partial T}\r_\epsilon=\l\frac{\partial n}{\partial T}\r + 
 \l\frac{\partial n}{\partial \mu}\r\l\frac{\partial \mu}{\partial T}\r_\epsilon$, 
 $\l\frac{\partial n}{\partial \mu}\r_\epsilon=\l\frac{\partial n}{\partial \mu}\r + 
 \l\frac{\partial n}{\partial T}\r\l\frac{\partial T}{\partial \mu}\r_\epsilon$.  We have 
$\l\frac{\partial T}{\partial \mu}\r_\epsilon= - 
 \frac{\l\frac{\partial \epsilon}{\partial \mu}\r}{\l\frac{\partial \epsilon}{\partial T}\r}$ 
along the constant $\epsilon$ trajectory  and  
finally 
 $\frac{\partial n}{\partial s}$ can be written as $\l\frac{\partial n}{\partial s}\r = 
 \l\frac{\partial n}{\partial T}\r\l\frac{\partial T}{\partial s}\r+
 \l\frac{\partial n}{\partial\mu}\r\l\frac{\partial\mu}{\partial s}\r$. Thus, bulk viscosity to entropy density 
 ratio in units of the relaxation time scale can be expressed as:
\beqa
\frac{\zeta}{s\tau_{\text{R}}} &=& -\l\frac{\partial P}{\partial n}\r_\epsilon
\l\frac{\partial n}{\partial s}\r\nn\\
&=&-\l\frac{\l \frac{\partial P}{\partial T} \r}{\l \frac{\partial n}{\partial T} \r 
- \l \frac{\partial n}{\partial \mu} \r\frac{\l\frac{\partial \epsilon}{\partial T}\r}{\l\frac{\partial 
\epsilon}{\partial \mu}\r}} + \frac{\l \frac{\partial P}{\partial \mu} \r}{\l \frac{\partial n}{\partial \mu} \r 
- \l \frac{\partial n}{\partial T} \r\frac{\l\frac{\partial \epsilon}{\partial \mu}\r}{\l\frac{\partial \epsilon}
{\partial T}\r}}\r\l \l\frac{\partial n}{\partial T}\r\l\frac{\partial T}{\partial s}\r+
 \l\frac{\partial n}{\partial\mu}\r\l\frac{\partial\mu}{\partial s}\r\r
\label{eq.app.zetabystau}
\eeqa
Now all the derivatives in Eq.~\ref{eq.app.zetabystau} can be evaluated starting from the expression of 
$\ln Z$ within the HRG model. To begin with, the partition function $Z\l T,V,\mu\r$ is given by 
\beq
\ln\, Z^{GC}(T,V,\{\mu_i\})=\sum_i \frac{g_i}{2\pi^2}VT^3\sum_{n=1}^\infty 
\frac{(\mp1)^{(n+1)}}{n^4}x_i^2 K_2(x_i)e^{y_i}\nn
\eeq
The pressure $P$ is obtained by operating $T\frac{\partial}{\partial V}$ on $ln Z^{GC}$
\beq
P^{GC}(T,V,\{\mu_i\})=\sum_i \frac{g_i}{2\pi^2}T^4\sum_{n=1}^\infty 
\frac{(\mp1)^{(n+1)}}{n^4}x_i^2 K_2(x_i)e^{y_i}\nn
\eeq
where $x_i=nm_i/T$ and $y_i=n\mu_i/T$ introduced for brevity in notation. 
Further the derivatives of $P$ are obtained as
\beqa
\left(\frac{\partial P^{GC}}{\partial T}\right) &=& \frac{1}{V}\left\{ln\, Z^{GC}+
\frac{1}{T}\left( E^{GC}-\sum_i \mu_i N_i^{GC}\right) \right\}\nn\\
\left(\frac{\partial P^{GC}}{\partial \mu_i}\right) &=& \frac{1}{V}\sum_i N_i^{GC}\nn
\eeqa
The particle number and its derivatives are given by
\beqa
N_i^{GC}(T,V,\mu_i) &=& T\frac{\partial \ln Z^{GC}}{\mu_i}\nn\\
N_i^{GC}(T,V,\mu_i) &=& \frac{g_i}{2\pi^2}VT^3\sum_{n=1}^\infty\frac{(\mp1)^{(n+1)}}{n^3}
x_i^2 K_2(x_i)e^{y_i}\nn\\
\left(\frac{\partial N_i^{GC}}{\partial T}\right) &=& \frac{g_i V m_i^2}{2\pi^2}\sum_{n=1}^\infty 
\frac{\l\mp1\r^{n+1}}{n}e^{y_i}\left[\frac{x_i}{2}K_1(x_i)+\l1-y_i\r
K_2(x_i)+\frac{x_i}{2}K_3(x_i)\right]\nn\\
\left(\frac{\partial N_i^{GC}}{\partial \mu}\right)_T &=& \frac{B_i g_i}{2\pi^2}VT^2
\sum_{n=1}^\infty\frac{(\mp1)^{(n+1)}}{n^2}x_i^2 K_2(x_i)e^{y_i}\nn
\eeqa

The energy and its derivatives are given by
\beq
E^{GC}(T,V,\{\mu_i\})=T^2\frac{\partial\ln Z^{GC}}{\partial T} + \sum_i \mu_iN_i^{GC}\nn\\
\eeq
\beqa
E^{GC}(T,V,\{\mu_i\}) &=& \sum_i {\cal C}_iT\sum_{n=1}^\infty \frac{(\mp1)^{(n+1)}}{n^2}
e^{y_i}\left[\frac{x_i}{2} K_1(x_i)\right.\nn\\
&&\left.+\l1-y_i\r K_2(x_i)
+\frac{x_i}{2} K_3(x_i)\right]
+\sum_i\mu_iN_i^{GC}\nn\\
\eeqa
\beqa
\left(\frac{\partial E^{GC}}{\partial T}\right)_\mu &=&\sum_i{\cal C}_i
\sum_{n=1}^{\infty}\frac{\l\mp1\r^{n+1}}{n^2}e^{y_i}\left[\frac{x_i^2}{4} K_0(x_i)
+(-x_iy_i+x_i)K_1(x_i)\right.\nn\\
&&\left.+(y_i^2+\frac{x_i^2}{2})+2(1-y_i)K_2(x_i)
+(-x_iy_i+x_i)K_3(x_i)
+\frac{x_i^2}{4} K_4(x_i)\right] +\sum_i\mu_i\frac{\partial N_i}{\partial T}
\eeqa
\beqa
\left(\frac{\partial E^{GC}}{\partial\mu}\right)_T &=&\sum_i{\cal C}_i\sum_{n=1}^\infty \frac{(\mp1)^{(n+1)}}{n}
e^{y_i}\left[\frac{x_i}{2}K_1(x_i) 
-y_i K_2(x_i)
+\frac{x_i}{2}K_3(x_i)\right]\nn\\
&&+\sum_i \l N_i+\mu_i\frac{\partial N_i}{\partial\mu}\r
\eeqa
where 
\beq
{\cal C}_i= g_i\frac{VT}{2\pi^2}m_i^2
\eeq
Finally the entropy and its derivatives are given by
\beqa
S^{GC}(T,V,\{\mu_i\}) &=& \frac{1}{T}\left\{E^{GC}(T,V,\{\mu_i\})+P^{GC}(T,V,\{\mu_i\})V
-\sum_i \mu_i N_i^{GC}(T,V,\mu_i)\right\}\nn\\
\left(\frac{\partial S^{GC}}{\partial T}\right) &=& -\frac{S}{T}+\frac{1}{T}\left\{\left(\frac{\partial E^{GC}}
{\partial T}\right)+V\frac{\partial P}{\partial T}-\sum_i \mu_i\left(\frac{\partial N_i^{GC}}{\partial T}\right) \right\}\nn\\
\left(\frac{\partial S^{GC}}{\partial \mu}\right) &=& \frac{1}{T}\left\{\left(\frac{\partial E^{GC}}
{\partial \mu}\right)+V\frac{\partial P}{\partial\mu}-\sum_iN_i-\sum_i \mu_i\left(\frac{\partial N_i^{GC}}{\partial \mu}\right) \right\}\nn
\eeqa
where we have used 
\beq
\frac{\partial}{\partial \mu}=\sum_i\frac{\partial \mu_i}{\partial \mu}\frac{\partial}{\partial \mu_i}\nn
\eeq
The speed of sound, $c_s$ can be used as a regulator for CSB. 
A system with $c_s^2\rightarrow 1/3$ will indicate the restoration of CSB. Therefore,
we the expression to estimate, $c_s$ within the ambit of present model is recalled below.
\beqa
c_s^2 &=& \l\frac{\partial P}{\partial \epsilon}\r_{s/n_B}\nn\\
&=& \l\frac{\partial P}{\partial T}\r\l\frac{\partial T}{\partial \epsilon}\r_{s/n_B} + 
\l\frac{\partial P}{\partial \mu}\r\l\frac{\partial \mu}{\partial \epsilon}\r_{s/n_B}\nn
\eeqa
Now for constant $s/n_B$, $d(s/n_B) = \frac{\partial (s/n_B)}{\partial T}dT + \frac{\partial (s/n_B)}{\partial \mu}d\mu=0$. 
This implies that $\l\frac{\partial T}{\partial \mu}\r_{s/n_B}=-\frac{\frac{\partial (s/n_B)}{\partial \mu}}
{\frac{\partial (s/n_B)}{\partial T}}$. Thus 
\beqa
\l\frac{\partial \epsilon}{\partial T}\r_{s/n_B} &=& \frac{\partial\epsilon}{\partial T}-\frac{\partial\epsilon}{\partial \mu}
\l\frac{\frac{\partial (s/n_B)}{\partial T}}{\frac{\partial (s/n_B)}{\partial \mu}}\r\nn\\
\l\frac{\partial \epsilon}{\partial \mu}\r_{s/n_B} &=& \frac{\partial\epsilon}{\partial \mu}-\frac{\partial\epsilon}{\partial T}
\l\frac{\frac{\partial (s/n_B)}{\partial \mu}}{\frac{\partial (s/n_B)}{\partial T}}\r\nn
\eeqa
Thus finally the expression for $c_s$ turns out to be
\beqa
c_s^2 &=& \l\frac{\l \frac{\partial P}{\partial T} \r_\mu}{\l \frac{\partial \epsilon}{\partial T} \r_\mu 
- \l \frac{\partial \epsilon}{\partial \mu} \r_T\frac{\l\frac{\partial (s/n_B)}{\partial T}\r}{\l\frac{\partial 
(s/n_B)}{\partial \mu}\r}} + \frac{\l \frac{\partial P}{\partial \mu} \r_T}{\l \frac{\partial \epsilon}{\partial \mu} \r_T 
- \l \frac{\partial \epsilon}{\partial T} \r_\mu\frac{\l\frac{\partial (s/n_B)}{\partial \mu}\r}{\l\frac{\partial (s/n_B)}
{\partial T}\r}}\r.\nn
\eeqa


\begin{thebibliography}{99}

\bibitem{Heinz:2013th} U.  Heinz  and S. Raimond, Ann. Rev. Nucl. Part. Sci.,
{\bf 63}, 123 (2013).

\bibitem{Gale:2013da} C. Gale, S. Jeon and B. Schenke, Int. J. Mod. Phys.
      A {\bf 28}, 1340011 (2013). 

\bibitem{Romatschke:2007mq} P. Romatschke and U. Romatschke, Phys. Rev. Lett.,
{\bf 99}, 172301 (2007).

\bibitem{Song:2012ua} H. Song, Nucl. Phys. A {\bf 904}, 114c (2013).

\bibitem{Alver:2010gr} B. Alver and G. Roland, Phys. Rev. C {\bf 81}, 054905 (2010).

\bibitem{Gardim:2011xv}
      F. G. Gardim,  F. Grassi, M. Luzum, J.-Y. Ollitrault,
      Phys. Rev. C {\bf 85}, 024908 (2012).

\bibitem{Niemi:2012aj} H. Niemi, G. S. Denicol, H. Holopainen and P.  Huovinen,
      Phys. Rev. C {\bf 87}, 054901 (2013).


\bibitem{dusling} K. Dusling and T. Sch\"afer,  Phys. Rev. C {\bf 85}, 044909 (2012).

\bibitem{Gyulassy:2004zy} M. Gyulassy and L. McLerran, Nucl. Phys. A {\bf 750}, 30 (2005).

\bibitem{Arsene:2004fa} I. Arsene {\it et al.} (for BRAHMS collaboration), 
Nucl. Phys. A {\bf 757}, 1 (2005). 

\bibitem{Adcox:2004mh} K. Adcox {\it et a.} (for PHENIX collaboration), 
      Nucl. Phys. A {\bf 757}, 184 (2005).yy

\bibitem{Back:2004je} B. B.  Back {\it et al.}  (for PHOBOS collaboration),
      Nucl. Phys. A {\bf 757}, 28 (2005).

\bibitem{Adams:2005dq} J. Adams {\it et al.} (for STAR collaboration)
      Nucl. Phys. A {\bf 757}, 102 (2005).

\bibitem{Kovtun:2004de} P. Kovtun, D. T. Son and A. O. Starinets,
Phys. Rev. Lett. {\bf 94}, 111601 (2005).
	
\bibitem{gubser} S. S. Gubser, A. Nellore, S. S. Pufu, F. D. Rocha 
Phys. Rev. Lett. {\bf 101}, 131601 (2008).

\bibitem{Meyer:2007dy} H. B. Meyer, Phys. Rev. Lett. {\bf 100}, 162001 (2008).


\bibitem{Karsch:2007jc} F. Karsch, D. Kharzeev and K. Tuchin, Phys. Lett. B {\bf 663},
217 (2008).


\bibitem{model1} A. Dobado and J. M. Torres-Rincon, Phys. Rev. {\bf D 86}, 074021 (2012). 

\bibitem{model2} S. Ghosh {\it et. al.}, Phys Rev {\bf C 93}, 045205 (2016).

\bibitem{model3} A. Tawfik, A. M. Diab and T. M. Hussein, Int. J. Adv. Research Phys. Science 
{\bf 3}, 4 (2016)

\bibitem{ryu} S. Ryu, J.-F. Paquet, C. Shen, G. S. Denicol, 
B. Schenke and C. Gale,  Phys. Rev. Lett. {\bf 115}, 132301 (2015).

\bibitem{Lu} E. Lu and G. D. Moore, Phys. Rev. C {\bf 83}, 044901 (2011).

\bibitem{sukanya} S. Mitra and S. Sarkar, Phys. Rev. D {\bf 87}, 094026 (2013).

\bibitem{juan} A . Dobado, F. J. Llanes-Estrada and J. M. Torres-Rincon, Phys. Lett. B {\bf 702}, 43 (2011).

\bibitem{guru} G. P. Kadam and H. Mishra, Phys. Rev. C {\bf 92}, 035203 (2015).

\bibitem{purnendu} P. Chakraborty and J. I. Kapusta, Phys. Rev. C {\bf 83}, 014906 (2011).

\bibitem{wang} J. -W. Chen and J. wang, Phys. Rev. C {\bf 79}, 044913 (2009).

\bibitem{sasaki} C. Sasaki and K. Redlich, Phys. Rev. C {\bf 79}, 055207 (2009).


\bibitem{noronha} J. Noronha-Hostler, J. Noronha and F. Grassi,
Phys. Rev. C {\bf 90}, 034907 (2014).


\bibitem{Paech:2006st} K. Paech and S. Pratt, Phys. Rev. C {\bf 74}, 014901 (2006).

\bibitem{denicolprc88} G. S. Denicol, C. Gale, S. Jeon and J. Noronha, Phys.
Rev. C {\bf 88}, 064901 (2013). 

\bibitem{Agashe:2014kda} K. A. Olive {\it et al.} (Particle Data Group), 
      Chin. Phys. C{\bf 38} 090001 (2014).


\bibitem{fernandez} D. Fernandez-Fraile and A. Gomez Nicola, 
Phys. Rev. Lett. {\bf 102}, 121601 (2009).

\bibitem{mcheng} M. Cheng {\it et al.}, Phys. Rev. D {\bf 77}, 014901 (2008). 

\bibitem{chatterjee} S. Chatterjee, R. M. Godbole and S. Gupta, Phys. Rev. {\bf C 81}, 044907 (2010).

\bibitem{Lo} P. M. Lo, M. Marczenko, K. Redlich and C. Sasaki,
Phys. Rev. C 92, 055206 (2015).

\bibitem{tawfikbulk}A. Tawfik and M. Wahba, Annalen Phys. {\bf522}, 849-856 (2010).

\bibitem{ejiri} S. Ejiri, F. Karsch, E. Laermann and C. Schmidt, Phys. Rev. D {\bf 73}, 054506 (2006).

\bibitem{tawfikcomments} A. Tawfik, E. Gamal, H. Magdy, arXiv:1301.1828 [nucl-th].

\bibitem{cleymans1} J. Cleymans and K. Redlich, Phys.
Rev. Lett. {\bf 81}, 5284 (1998).

\bibitem{cleymans2} J. Cleymans, H. Oeschler, K. Redlich, and S. Wheaton, Phys.
Rev. C {\bf 73}, 034905 (2006).

\bibitem{LQCD-data} S. Borsányi {\it et. al.}, Physics Letters {\bf B 730}, 99 (2014).

\bibitem{goren} V. Vovchenko, D.V. Anchishkin and M.I. Gorenstein, Phys. Rev. {\bf C 91}, 024905 (2015).


\end{thebibliography}
\end{document}